\documentclass[pra,superscriptaddress,showpacs,papersize=a4paper,twocolumn,floatfix]{revtex4-1}
\usepackage[pdftex]{graphicx}
\usepackage{bm}
\usepackage{graphicx,subfigure}
\usepackage{etoolbox}
\usepackage[usenames,dvipsnames]{color}
\usepackage{mathtools}
\usepackage{amsmath}
\usepackage{amsfonts}
\usepackage{amssymb}

\DeclarePairedDelimiterX\braket[2]{\langle}{\rangle}{#1 \delimsize\vert #2}

\newcommand{\bg}{ \begin{gather} }
\newcommand{\eg}{\end{gather}}
\newcommand{\be}{ \begin{equation} }
\newcommand{\ee}{\end{equation}}
\newcommand{\bea}{ \begin{eqnarray} }
\newcommand{\eea}{\end{eqnarray}}

\begin{document}

\title{Many-body localization transition with power-law interactions: Statistics of eigenstates}

\author{K.\,S.~Tikhonov}
\affiliation{Institut f{\"u}r Nanotechnologie, Karlsruhe Institute of Technology, 76021 Karlsruhe, Germany}
\affiliation{L.\,D.~Landau Institute for Theoretical Physics RAS, 119334 Moscow, Russia}

\author{A.\,D.~Mirlin}
\affiliation{Institut f{\"u}r Nanotechnologie, Karlsruhe Institute of Technology, 76021 Karlsruhe, Germany}
\affiliation{Institut f{\"u}r Theorie der Kondenserten Materie, Karlsruhe Institute of Technology, 76128 Karlsruhe, Germany}
\affiliation{L.\,D.~Landau Institute for Theoretical Physics RAS, 119334 Moscow, Russia}
\affiliation{Petersburg Nuclear Physics Institute,188300 St.\,Petersburg, Russia.}

\begin{abstract}
We study spectral and wavefunction statistics for many-body localization transition in systems with long-range interactions decaying as $1/r^\alpha$ with an exponent $\alpha$ satisfying $ d \le \alpha \le 2d$, where $d$ is the spatial dimensionality. We refine earlier arguments and show that the system undergoes a localization transition as a function of the rescaled disorder $W^* = W / L^{2d-\alpha} \ln L$, where $W$ is the disorder strength and $L$ the system size. This transition has much in common with that on random regular graphs. We further perform a detailed analysis of the inverse participation ratio (IPR) of many-body wavefunctions, exploring how ergodic behavior in the delocalized phase switches to fractal one at the critical point and on the localized side of the transition. Our analytical results for the scaling of the critical disorder $W$ with the system size $L$ and for the scaling of IPR in the delocalized and localized phases are supported and corroborated by exact diagonalization of spin chains. 
\end{abstract}
\maketitle


\section{Introduction}
\label{s1}

Many-body localization (MBL) has recently become a powerful idea in the theory of disordered interacting quantum systems. 
The MBL extends the Anderson-localization physics originally formulated for a single-particle problem \cite{anderson1958absence} to many-body systems at non-zero energy density (or, equivalently, non-zero temperature). 
Starting from early works \cite{fleishman1980interactions,altshuler1997quasiparticle,gornyi2005interacting,basko2006metal,oganesyan2007localization} and until recently, understanding of MBL was driven mostly by theory. In particular, Refs. \cite{gornyi2005interacting,basko2006metal} predicted a finite-temperature MBL transition for spatially extended systems with localized single-particle states and with short-range interaction. This result has been supported and refined by numerous subsequent numerical and analytical studies, see, in particular, Refs.~\onlinecite{oganesyan2007localization,monthus10,kjall14,gopalakrishnan14,luitz2015many,karrasch15,imbrie16,imbrie16a,gornyi2017spectral} as well as reviews \cite{nandkishore2015many,luitz2017ergodic,abanin2017recent}.

Recently, experimental realizations of one-dimensional (1D) \cite{schreiber2015observation,smith2016many} and two-dimensional (2D) \cite{kondov2015disorder,choi2016exploring} systems showing MBL transition were implemented for cold atoms in disordered optical lattices. Signatures of MBL transition in interacting systems were also observed  in InO films~\cite{ovadyahu1,ovadyahu2,ovadia2015evidence}. Further, the MBL was studied experimentally in arrays of coupled one-dimensional optical lattices~\cite{bordia15,lueschen16}, spin impurities in diamond \cite{choi2017observation}, and atomic ions \cite{Zhang17}. Spectroscopic signatures of MBL were also observed in systems of coupled superconductinbg qubits \cite{roushan17}.

While Refs. \cite{gornyi2005interacting,basko2006metal} dealt with systems with short-range interaction, in many of experimentally relevant systems interactions are in fact long-ranged in the sense that they decay with distance $r$ according to a power law. Consider, for example, electrons in an Anderson insulator such as a 2D system on the quantum Hall plateau. In the absence of long-range interactions, a bulk of such a system would be in the MBL phase at low temperatures $T$. However, it has been found experimentally that there is quite essential energy transport through the bulk of
integer  \cite{granger2009observation,venkatachalam2012local} and fractional  \cite{altimiras2012chargeless, inoue2014proliferation} quantum Hall systems. On the theory level, such heat transport becomes possible due to dipole-dipole coupling between two-level systems (``spins'') formed by nearby localized electronic states. This power-law $1/r^3$  interaction of ``spins'' (originating from the $1/r$ Coulomb interaction between electrons) leads to many-body delocalization, establishing a finite thermal conductivity that has a power-law dependence on temperature at low $T$ \cite{gutman2015energy}. This example demonstrates the importance of understanding of the physics of many-body-localization and -delocalization in systems with long-range interactions. In addition to electronic realizations, the problem of (de-)localization in disordered many-body systems with dipole interactions arises also in other contexts, including amorphous materials (glasses)  \cite{Zeller71,Hunklinger86} that can be described in terms of interacting two-level systems \cite{Anderson72,Phillips72,Yu88,Burin89,Burin1998,Burin94,Classen00,Burin04},  dipolar molecules in an optical lattice   \cite{Barnett06,Gorshkov11,yao2014many,Yan13,Hazzard14}, spin defects in a solid-state system \cite{yao2014many,Choi17a,Choi16,choi2017observation,Ho17}, as well as superconducting circuits \cite{Lisenfeld10a,Burnett14,Lisenfeld10b}. 
Further, an experimental realization of a one-dimensional system of trapped ions with tunable long-range interaction that can be approximated by a power law with a tunable exponent has been reported in Refs.~\onlinecite{Monroe13,smith2016many}. 

Theoretical investigation of the effect of long-range terms on localization has in fact a long history.
For a non-interacting problem with strong disorder and hopping terms decaying as $r^{-\alpha}$ in a spatial dimensionality $d$, it was shown already in the famous  Anderson's paper in 1958 \cite{anderson1958absence} that at $\alpha<d$ the locator expansion breaks down due to a diverging number of resonances.  This conclusion was confirmed by later works where the power-law-hopping non-interacting problem was analyzed in much detail, see, in particular, Refs.\cite{levitov1989absence,levitov1990delocalization,mirlin1996transition,levitov1999critical,Mirlin00}. For a problem with a long-range interaction, 
considering the effect of interaction in the first order, one gets an approximate mapping to the non-interacting problem \cite{levitov1999critical}. It turns out, however, that this argument is too naive. Specifically, a more efficient delocalization mechanism has been identified, implying absence of localization in the thermodynamic limit for an arbitrarily strong disorder already for $\alpha < 2d$ \cite{burin1994low,burin2006energy,yao2014many,burin2015many,gutman2015energy}. Thus, a new phase arises at $d \le \alpha < 2d$
that would be localized (for $\alpha > d$) or critical (for $\alpha = d$) within an approximate mapping to a non-interacting power-law  problem but is in reality many-body-delocalized in the thermodynamic limit \cite{footnote-confinement}. 

While a system with long-range interaction exponent satisfying $d \le \alpha < 2d$ is delocalized in the thermodynamic limit, its finite-size properties are by no means trivial. Specifically, such a system exhibits a many-body delocalization transition with increasing size $L$ \cite{burin2015many,gutman2015energy}.  The goal of this work is to explore the position of this transition as well as the spectral and eigenfunction statistics at and around the transition. We will put particular emphasis on the statistics of many-body wavefunctions with varying disorder and system size.  

One of the approaches to the theory of MBL  is based on approximate mapping of an interacting Hamiltonian to a non-interacting hopping problem defined on a hierarchical lattice. This idea was first put forward in Ref.~\cite{altshuler1997quasiparticle} in the context of a hot-electron relaxation in a quantum dot and later employed in a number of papers for the analysis of the MBL transitions. This connection with the localization in many-body systems has recently revived an interest to the problem of Anderson localization of non-interacting fermions residing on tree-like lattices such as random regular graphs (RRG) and their close relatives \cite{Biroli12,DeLuca14,tikhonov2016anderson,garcia-mata17,metz17}. This problem was in fact studied analytically via supersymmetry method long ago \cite{mirlin1991universality,fyodorov1991localization,fyodorov1992novel}. The corresponding analytical predictions for the level and eigenfunctions statistics near the transition have been supported and corroborated by recent numerical works \cite{tikhonov2016anderson,garcia-mata17,metz17}. 
In the present paper, we analyze the connection between many-body and RRG problems. We show that the MBL transition in a many-body problem with a long-range interaction with $d<\alpha<2d$ is particularly close to the Anderson transition on RRG.  
Combining analytical considerations and exact-diagonalization numerics, we perform a detailed  study of the statistics of energy levels and eigenfunctions that allows us to establish the scaling of the MBL transition in the power-law-interaction problem and to explore properties of the system at the critical point and around it. We show, in particular, that the critical point essentially shares properties of the localized phase, including the Poisson statistics (in the limit of large $L$) and the fractal scaling of the inverse participation ratio (IPR) with the Hilbert-space volume. On the other hand, on the delocalized side of the transition, the system becomes ergodic in the large-$L$ limit.

We consider a system of spins 1/2 described by the following Hamiltonian:
\begin{equation}
\label{long_spin}
\hat H = \sum_i\epsilon_i \hat \sigma_i^z+t\sum_{ij}\frac{u_{ij}\hat \sigma_i^z \hat \sigma_j^z+v_{ij}(\hat \sigma_i^+ \hat \sigma_j^-+\hat \sigma_i^- \hat \sigma_j^+)}{r_{ij}^{\alpha}},
\end{equation}
with independent random variables $u_{ij}, \; v_{ij}=\pm 1$ and with $\epsilon_i$ sampled uniformly from the interval $\left[-W/2,W/2\right]$. 
Here $\hat\sigma_i^z$, $\hat\sigma_i^+$, and  $\hat\sigma_i^-$ are Pauli matrices and $r_{ij}$ is the distance between the sites $i$ and $j$.
Analytically, we consider a $d$-dimensional version of this Hamiltonian; in numerical simulations, we study 1D lattices of $L$ spins  via exact diagonalization.  

The structure of the article is as follows. We first recall in Sec.~\ref{s2} mechanisms leading to many-body delocalization of the system Eq. (\ref{long_spin}) for sufficiently long-ranged interactions, $\alpha<2d$, in the thermodynamic limit of $L\to\infty$. 
Then, in Sec.~\ref{s3}, we turn to the connection of this model with the Anderson model on RRG. This yields, in particular, the scaling of the critical disorder $W_c(L)$ with the system size $L$.  Using the many-body level statistics, we provide a numerical evidence for this mapping between the MBL transition in a power-law-interaction model and the Anderson transition on RRG and determine a position of the transition. In Sec.~\ref{sec:WF} we analyze the structure of many-body wavefunctions and explore the scaling of the corresponding IPR in localized and delocalized phases as well as at criticality. 
These analytical estimates are in good agreement with numerical data obtained from exact diagonalization. A numerical analysis of IPR provides an alternative method of determination of the position of the MBL transition, yielding results that are fully consistent with those obtained from the spectral statistics. In Sec.~\ref{sec:phase-diag} we discuss the width of the critical regime around $W_c$ that separates the localized and delocalized phases. We conclude the paper by summarizing our results and discussing prospects for future research in Sec.~\ref{s5}.

\section{Mechanisms of delocalization and critical dimensionality}
\label{s2}

Let us first discuss the noninteracting counterpart of the model (\ref{long_spin}) which describes a particle hopping over a $d$-dimensional lattice with random hopping amplitude decaying as a power-law $1/r^\alpha$ with the distance $r$. It is known \cite{anderson1958absence,levitov1989absence,levitov1990delocalization} that the point $\alpha=d$ is critical for this model.
For $\alpha<d$ the single-particle excitations in such a model delocalize at arbitrary disorder strength due to a diverging number of resonances. For larger power-law exponents, $\alpha> d$, the Anderson localization becomes possible in the thermodynamical limit. 
A particularly detailed analytical and numerical study has been performed for a 1D model of this class, known as power-law random banded matrix (PRBM) ensemble \cite{mirlin1996transition,Mirlin00}. In this model, the random  off-diagonal (i.e., hopping) matrix elements are characterized by a variance that 
decays as $b/|i-j|^{2\alpha}$ for $|i-j| > b$, while the diagonal matrix elements have a variance unity. It was found that all eigenvectors are localized for $\alpha > 1$ and delocalized for $\alpha < 1$. For $\alpha=1$ the energy levels and eigenfunctions statistics are critical for any value of $b$.

Now let us return to the interacting model (\ref{long_spin}). Localization in a system with long-range interaction was first discussed in Ref.~\onlinecite{fleishman1980interactions}, where it was argued, by analogy with Ref.~\cite{anderson1958absence},  that the interaction delocalizes the system at $\alpha<d$. This argument may seem to suggest that the critical point $\alpha=d$ of a non-interacting model is also critical for the interacting model \cite{levitov1999critical}. It turns out, however, that the critical dimensionality of the interacting problem (\ref{long_spin}) is in fact  lower, $d_c=\alpha/2$ \cite{burin1994low,burin2006energy,yao2014many,gutman2015energy}. In other words,  the power-law-interaction problem exhibits a more efficient mechanism of delocalization than the non-interacting power-law-hopping model. This delocalization mechanism originates from resonant interactions between resonant spin pairs. 
Below we briefly reiterate the corresponding arguments, which will also play an important role for a later discussion of the statistics of the many-body wavefunctions.

We consider a regime of strong disorder, so that a starting point is a basis of many-body states with all spins having definite $z$ components $\sigma_i^z = \pm 1$.
Each spin in such a state has an energy $\bar{\epsilon}$ renormalized due to interaction with other spins:
\be
\bar{\epsilon_i}=\epsilon_i+t\sum_k r^{-\alpha}_{ik}u_{ik}\sigma_i^z\sigma_k^z.
\label{bar-epsilon}
\ee
Two spins $i$ and $j$ are in resonance if 
\be
\left|\bar{\epsilon_i}-\bar{\epsilon_j}\right|\lesssim\frac{t}{r_{ij}^{\alpha}}.
\ee 
Two strongly hybridized levels of such a resonant pair (those with total $z$ projection of spin equal to zero) form a new degree of freedom, \emph{pseudospin}. For a given spin, an average number of its resonance partners within a layer $R<|r_i-r_j|<2R$ is estimated as
\be
N_{1}(R)=t\rho R^{d-\alpha}/W.
\ee
The spatial density of ``active'' pseudospins (resonant pairs build out of original spins with energies within the thermal window) of size $\sim R$ is thus
\be
\rho_{\rm PS}(R)=\rho N_{1}(R)T/W\sim t\rho^2 T R^{d-\alpha}/W^2.
\label{rho-ps}
\ee
It is assumed in Eq.~(\ref{rho-ps}) that the temperature $T$ is smaller than the bandwidth $W$. In the opposite case, the factor $T/W$ should be replaced by unity. 
For a sufficiently long-ranged interaction, $\alpha<d$, the density of pseudospins increases with $R$, which clearly implies delocalization. This is essentially the mechanism of Ref.~\cite{anderson1958absence}. 

We are interested, however, in the case of faster decaying interaction, $\alpha > d$, when $\rho_{\rm PS}(R)$ decreases with increasing $R$, so that most of the pseudospins have a microscopic size. Since we also assume a strong-disorder regime, such resonances are relatively rare, i.e., most spins do not have any resonant partner. 

Interestingly, although the density of pseudospins is low, their total number in the system may be much larger than unity even in the localized phase. 
However, in the localized phase such pseudospins typically ``do not talk to each other'' and for this reason do not induce delocalization. 
On the other hand, they do manifest themselves in the scaling of IPR, as will be discussed below in Sec.~\ref{sec:WF}.  

Now we turn to the discussion of the mechanism for delocalization based on interaction of the pseudospins.
For simplicity, we focus in the rest of the paper on the limit of infinite temperature (which effectively means $T\gtrsim W$), in which case the density of pseudospins takes the form
\be 
\rho_{\rm PS}(R)=t\rho^2 R^{d-\alpha}/W.
\ee
The number of pseudospins of size $\sim R$ within a volume of the linear size $\sim R$ is thus
\be
\label{NR}
N_2(R)\sim \frac{t\rho^2}{W}R^{2d-\alpha}.
\ee
For $\alpha<2d$, the function $N_2(R)$ monotonically increases with $R$. 
Let us now consider a finite system of linear size $L$. When the system is sufficiently small, we have $N_2(L) \ll 1$, so that there are no pseudospins of size $\sim L$. The existent pseudospins have much smaller size and 
do not ``communicate'' with each other. As a result, the system is in the localized regime. On the other hand, with increasing system size, $N_2(L)$ increases and eventually becomes larger than unity. This means that there are multiple pseudospins of size $L$ in the system. As we discuss in more detail in the next section, the interaction between pseudospins leads to many-body delocalization of the whole system. Already at this stage, we can anticipate that there is a line in the $W$--$L$ plane where delocalization takes place. In other words, the critical disorder $W_c$ depends non-trivially on the system size $L$. 

This means that the mere definition of the localization transition point in the  large-$L$ limit requires proper scaling of the disorder $W$ with the system size. Below we
analyze this scaling and then study properties of the spectrum and of many-body eigenstates around the transition. 

\section{Many-body localization transition}
\label{s3}

\begin{figure}[h!]
\includegraphics[width=0.95\linewidth]{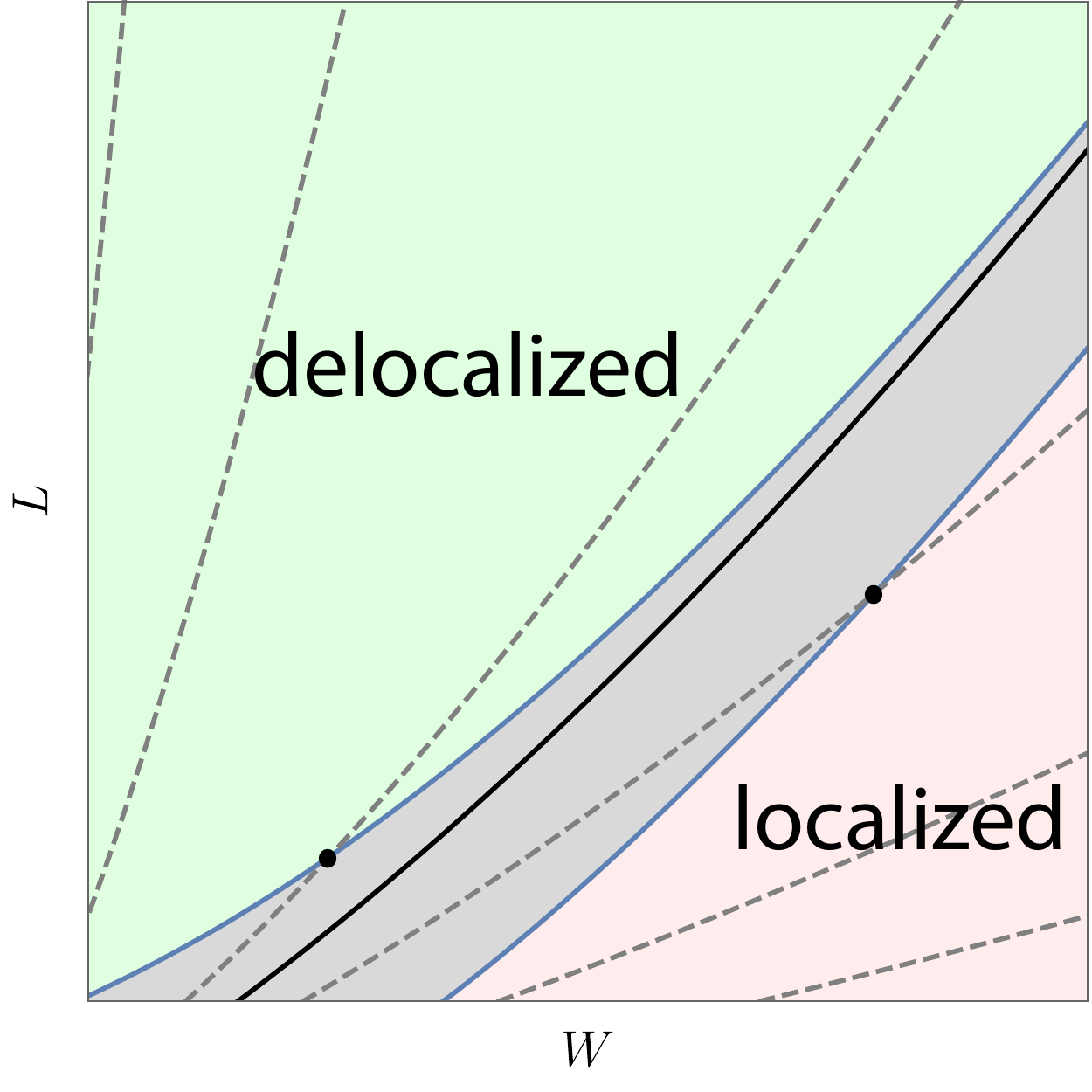}
\caption{Schematic phase diagram in the $W$--$L$ plane. The thick line is the critical line of the MBL transition, $W= W_c(L)$, or, equivalently, $W_* = W_{*c}$. 
The localized phase, the critical regime, and the delocalized phase are shown by pink, gray and green colors, respectively. 
Dashed lines correspond to fixed values of the renormalized disorder $W_*$.   Borders of the critical regime are determined by Eq.~(\ref{crit-regime}), where the correlation lengths  $\zeta(W_*)$ are given by Eq.~(\ref{corr-length}). For two values of $W_*$ (one slightly above $W_{*c}$ and another one slightly below),  black dots mark length scales where the system leaves the critical regime entering localized or, respectively, delocalized phase.}
\label{fig:phase-diagram}
\end{figure}
\subsection{Scaling of the critical point}
\label{sec:scaling}

As has been argued above, for a given disorder $W$, the system experiences a transition to the delocalized phase with increasing system size $L$. 
Equivalently, a system of given size $L$, undergoes a transition to the localized phase with increasing disorder $W$. 
In order to determine the corresponding critical length $L_c(W)$, or equivalently, the critical disorder $W_c(L)$, we begin with the following estimate.
Let us find the system size at which the pseudospins that are in resonance with each other start to emerge. This is found by setting $N_2(L) \sim 1$, with $N_2(L)$ given by Eq.~(\ref{NR}), which yields
\be
L_{c1}(W) \sim \left(W/t\rho^2\right)^{\frac{1}{2d-\alpha}},
\label{LW}
\ee
or, equivalently,
\be
W_{c1}(L)  \sim t\rho^2 L^{2d-\alpha}.
\label{WL}
\ee
This scaling of the critical point was proposed in Refs.~\cite{burin2015many,gutman2015energy}. The identification of Eq. (\ref{WL}) as the critical disorder of the MBL transition, however, is not at all trivial, as we are now going to discuss. Moreover, we will argue that Eq. (\ref{WL}) is not fully correct in the sense that it misses a logarithmic correction to scaling. 

Consider a system of size $L$ of the order of a few (order unity) $L_{c1}(W)$. A typical product state of this system at infinite temperature has a few pseudospins of size $\sim L$, i.e., it is well coupled to several other many-body states by the corresponding spin-flip interaction matrix elements Flipping any of the pseudospins provides another many-body state well connected with the original one. The new state will again have a few pseudospins and this process can be iterated. The crucial question is whether the new many-body states will have resonances \emph{distinct} from those  encountered on the previous steps of this iteration procedure. 
If we would discard $zz$ interactions, this would not be the case, and we would get stuck after a few steps. However, the $zz$ interactions that shift the energy 
$\bar{\epsilon_i}$ 
of a spin when other spins are flipped, see Eq.~(\ref{bar-epsilon}), are of crucial importance here. This effect is known as \emph{spectral diffusion}. As a result of these energy shifts, existing resonances are eliminated and new resonances are created. Specifically, after $p$ spin-flip steps, a typical distance to the closest flipped spin will be $\sim Lp^{-1/d}$, so that the typical shift of the energy 
$\bar{\epsilon_i}$  is estimated as \cite{gutman2015energy}
\be 
\Delta^{(p)}\bar{\epsilon_i} \sim tL^{-\alpha}p^{\alpha/d}.
\label{Delta-epsilon}
\ee

This fast increase of $\Delta^{(p)}$ with $p$  (we recall that $\alpha > d$) ensures that the resonances are very efficiently ``reshuffled'', so that the emerging network of many-body states coupled by such resonances has a tree-like structure \cite{gutman2015energy}. While locally this structure reminds a Cayley tree, the many-body Hilbert space is finite and has no boundary. Therefore, the resonant structure emerging in the many-body Hilbert space may be viewed as a RRG. 
We thus conclude that systems of sizes larger than $L_{c1}(W)$ should be ergodic.
Indeed, exact-diagonalization results in Ref.~\cite{burin2015many} supported this expectation. 

Let us note in passing that the spectral diffusion is also relevant for MBL transition in systems with short-range interactions \cite{gornyi2017spectral}, where it shifts the MBL transition point (parametrically enhancing delocalization) with respect to earlier estimates \cite{gornyi2005interacting,basko2006metal}. The spectral diffusion in that case (and in the case of quantum dot models) is somewhat less efficient, however, which made the analysis in Ref.~\cite{gornyi2017spectral} substantially more complicated.

Thus, the spectral diffusion ensures that systems with disorder weaker than $W_{c1}(L)$ are ergodic. Is a system with $W$ larger than $W_{c1}(L)$ necessarily localized? The answer is no. Indeed, the situation we are facing here is similar to the one known from Refs.~\onlinecite{anderson1958absence,abou1973selfconsistent} where it was shown that estimate based on counting of ``real'' first-order resonances underestimates the critical disorder of Anderson localization on a lattice with connectivity $K \gg 1$ by a logarithmic factor $\ln K$. This enhancement of delocalization arises from higher-order
resonances, i.e., those attainable via intermediate out-of-resonance states.
Such resonances lead to enhancement of the effective matrix element in $p$-th order of the perturbation theory by a factor $(\ln K)^{p-1}$, which yields, in the large-$p$ limit, the enhancement of $W_c$ by a factor $\sim\ln K$. More specifically, for the box distribution of disorder, the hopping matrix element set to unity, and in the middle of the band, the critical disorder is given by 
\be
W_c / K \simeq 4 \ln K.
\label{abou}
\ee
\begin{figure*}
\centering
\includegraphics[width=0.48\textwidth]{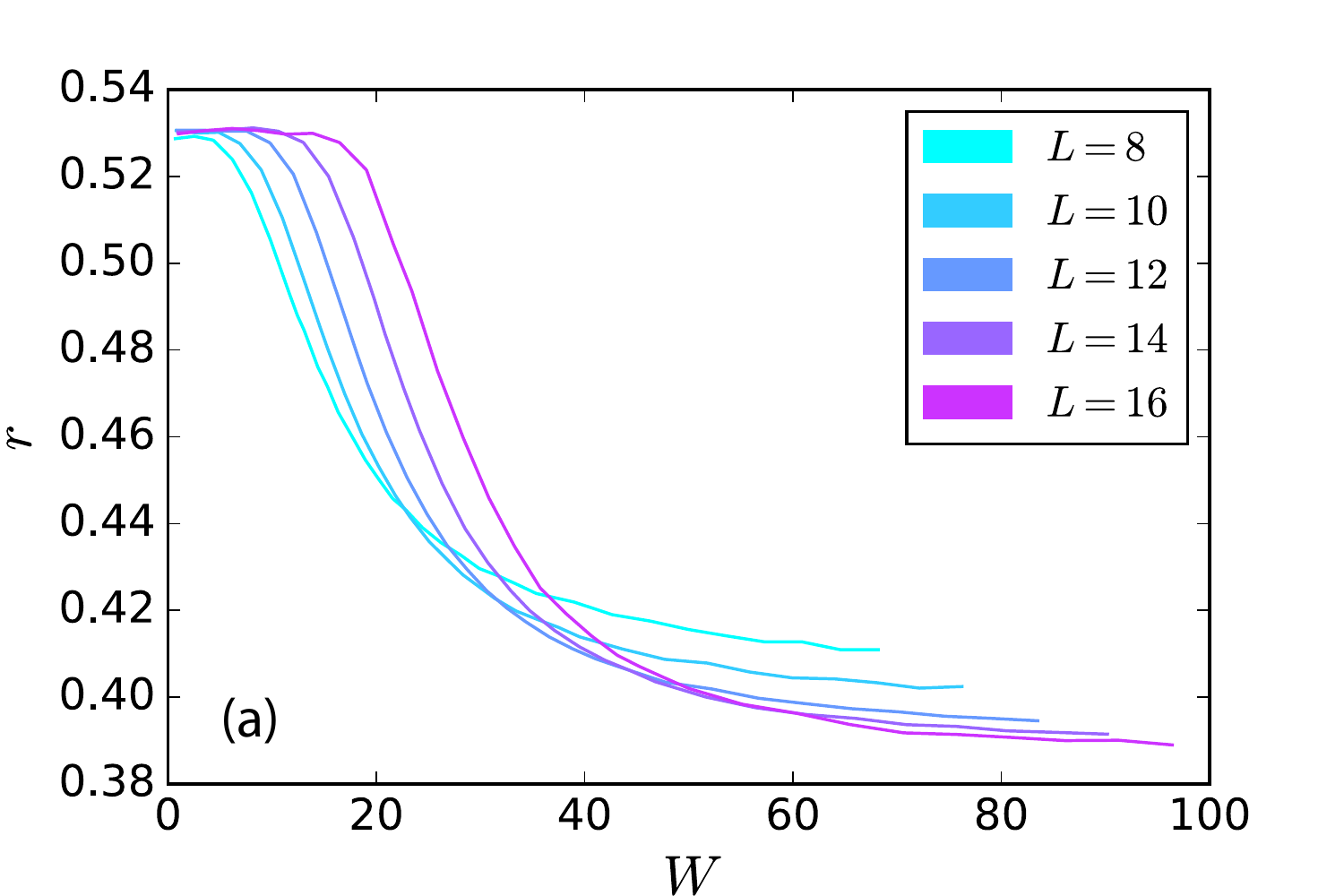}\quad
\includegraphics[width=0.48\textwidth]{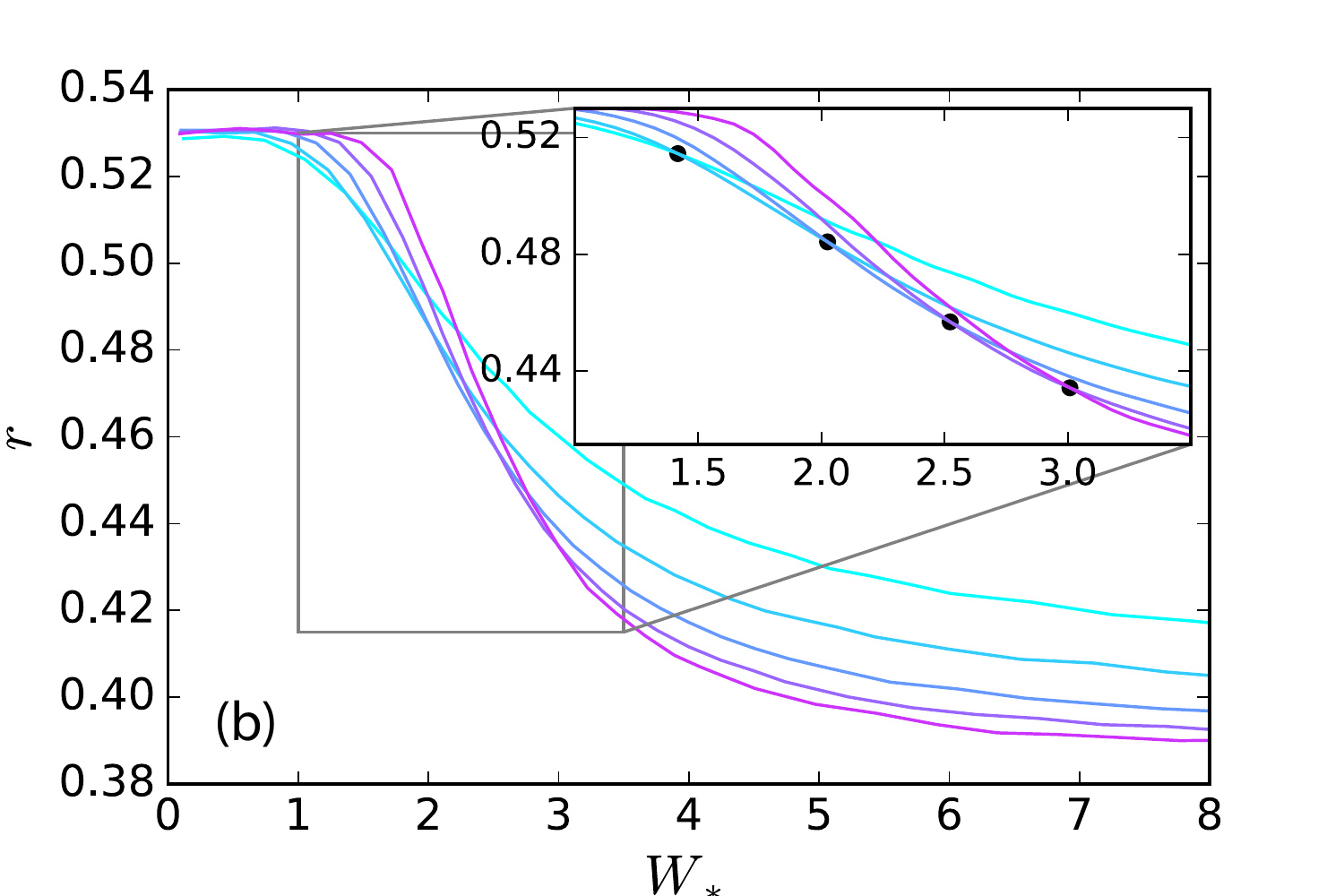}
\caption{Spectral statistics $r$ of a 1D spin chain (\ref{long_spin}) with $\alpha=3/2$ as a function of disorder $W$ for various system sizes $L$.
(a)  $r(W)$ demonstrating delocalization (ergodicity, $r \to r_{\rm WD}$) at fixed $W$ in the limit $L\to\infty$;
(b)  $r(W_*)$ with disorder rescaled according to Eq.~(\ref{W-star}). A drifting crossing point is observed that is expected to converge to a critical value $W_{*c}$ at $L \to \infty$, see Fig.~\ref{fig:crossing}. In the thermodynamic limit, $L \to \infty$, the system is expected to be ergodic for $W_* < W_{*c}$ and localized for $W_* > W_{*c}$. 
\label{fig:r}
}
\end{figure*}
We thus have to identify what plays the role of $\ln K$ in our problem.  The total connectivity of the graph obtained by counting all many-body basis states coupled to a given one by an interaction term is $K_{\rm tot} \sim \rho^2 L^{2d}$. In fact, one should exert a certain care here, since not all these states necessarily contribute to the logarithmic enhancement. Indeed, an amplitude of a higher-order process in a many-body system may be suppressed, in comparison with that on a Cayley-tree model, due to partial cancellations between the processes with permutations of elementary interaction processes. However, the spectral diffusion interferes also at this point, ensuring that the logarithmic enhancement is operative, since the shifts $\Delta^{(p)}\bar{\epsilon_i}$, Eq.~(\ref{Delta-epsilon}) destroy the cancellation \cite{gornyi2017spectral}. Specifically, consider a typical order of the perturbation theory in which we can rich any state starting from the given one, $p \sim L^d$. The corresponding energy shift is then given by interaction between nearby spins, $\Delta^{(p)} \bar{\epsilon_i} \sim t \rho^{\alpha/d}$. Thus, we expect a logarithmic factor originating from energy interval between  $\Delta^{(p)} \bar{\epsilon_i}$ and the spacing $W/ \rho^2 L^{2d} \sim t/L^{\alpha}$. This yields the factor $\ln K_{\rm eff}$ with $K_{\rm eff} \sim \rho^{\alpha/d} L^\alpha$. We see that the difference between  $\ln K_{\rm eff}$ and $\ln K_{\rm tot}$ is just a factor $\alpha/2d$ of order unity. To summarize, we predict a MBL transition at 
\begin{equation}
W_c(L)\sim t\rho^2 L^{2d-\alpha} \ln (\rho L^d),
\label{scaling2}
\end{equation}
up to a numerical coefficient of order unity. 

Equation (\ref{scaling2}) is different from Eq.~(\ref{WL}) by a logarithmic factor that was not taken into account in Refs.~\cite{burin2015many,gutman2015energy}. Of course, in the large-$L$ limit, this logarithm is not too important in comparison with the dominant power-law factor.  On the other, for system sizes $L$ than can be achieved in numerical simulations (see below) and in experiments, the logarithmic factor plays quite an important role. 

As usual, for finite $L$, the true localization transition turns to a crossover. The transition emerges, strictly speaking, in the thermodynamic limit, $L\to \infty$. The specialty of the present problem is that the thermodynamic limit should be taken by sending simultaneously $L$ and $W$ to infinity and keeping the ratio $W/W_c(L)$ fixed.  Then, for $W/W_c(L) > 1$ we will be in the MBL phase, for $W/W_c(L) < 1$ in the many-body delocalized phase, and for $W/W_c(L) = 1$ in the MBL transition critical point. The phase diagram in the plane spanned by $W$ and $L$ is illustrated in Fig.~\ref{fig:phase-diagram}. 

In view of the relation to the localization transition on RRG established above, we expect that the MBL transition in the present problem has the same gross features as the Anderson transition on RRG. Specifically, this implies that, in the large-$L$ limit taken as explained above,
(i) the level statistics at the critical point point is of Poisson form, as in the localized phase and (ii) the delocalized phase is ergodic   in the sense of Wigner-Dyson (WD) level statistics and of the scaling of the many-body IPR
\cite{mirlin1991universality,fyodorov1991localization,fyodorov1992novel,tikhonov2016anderson,garcia-mata17,metz17}. 

Below we will verify and supplement these predictions  by exact diagonalization of a 1D model described by Eq. (\ref{long_spin}) with $\alpha=3/2$. Specifically, we will first study the level statistics and then turn to the many-body eigenfunction statistics (IPR).  It is worth mentioning already here that there is a difference between the present model and RRG with respect to scaling of IPR in the localized phase that is related to short-scale resonances (that play no role for the transition mechanism) mentioned above. We will discuss implications of these resonances for the structure of many-body wave functions in Sec.~\ref{sec:WF}.

\subsection{Numerical analysis: Spectral statistics}
\label{sec:num-spectral}

In the numerical analysis, we consider spins on a regular 1D lattice with unit spacing (i.e., set the spin density to be $\rho = 1$), so that system size $L$ represents also the number of spins, ranging from $L=8$ to $L=16$. We consider periodic boundary conditions and set $t=1$ in Eq.~(\ref{long_spin}).

We consider the sector of vanishing total $\sigma_z$ and concentrate on $1/8$ of the states in the middle of the many-body band (which corresponds to taking the infinite temperature). 
In order to be able to generate large statistical ensembles for a wide range of $W$ (which is of crucial importance for a reliable analysis of the data), we restrict ourselves in the numerical analysis by values of the system size $L$ up to 16. This size is somewhat smaller than typical values of $L$ for exact-diagonalization studies of systems with short-range interactions, since our Hamiltonian matrix is by far less sparse than those in the short-range-interaction case. On the other hand, exponentially rare events are not important for the localization-delocalization transition and the physics around it in our problem, at variance with short-range-interaction problems where such events were argued to be essential (see also a discussion in the end of Sec.~\ref{s5}). This is favorable for the numerical analysis of the transition based on results for relatively small systems in the power-law-interaction model.

Disorder averaging was performed over $10^6$ realizations (smallest systems, $L=8$) to  $2\cdot 10^3$ realizations (largest systems $L=16$) at each $W$.  As a convenient scaling variable characterizing the spectral statistics, we use the ensemble-averaged ratio $r=\langle{r_i}\rangle$ of two consecutive spacings,  $r_i = \min(\delta_i,\delta_{i+1})/\max(\delta_i,\delta_{i+1})$, which takes values between $r_{\rm{P}}=0.386$ and $r_{\rm{WD}}=0.530$ realized for the Poisson and the WD Gaussian orthogonal ensemble (GOE) limits, respectively. The results for $\alpha=3/2$ are shown in the Fig. \ref{fig:r}a. The rapid shift of the curve $r(W)$ to the right with increasing $L$ fully supports the analytical expectation that for fixed $W$ the system is delocalized in the large-$L$ limit. Indeed, as Fig. \ref{fig:r}a indicates, for a fixed disorder $W$ the parameter $r$ approaches its ergodic value $0.530$ at $L\to\infty$. 

Let us now rescale the disorder according to the predicted scaling Eq. (\ref{scaling2}). Specifically, we define 
\be 
W_* = \frac{W}{L^{2d-\alpha} \ln L}.
\label{W-star}
\ee
The result is shown in  Fig. \ref{fig:r}b. The curves show now a behavior similar to the one found for Anderson model on RRG \cite{tikhonov2016anderson}. They get steeper with increasing $L$ and show a crossing point between curves corresponding to $L$ and $L+2$ total spins.  The zoomed-in region illustrates that this crossing point drifts to the right and that this drift is slowing down with increasing $L$, see also Fig.~\ref{fig:crossing}.  (Taking into account the logarithmic factor in the denominator of Eq.~(\ref{W-star}) is essential for this analysis; if this factor is discarded, the drift of the crossing point accelerates,  indicating a divergence in the large-$L$ limit.)
These results support the analytical prediction of the scaling (\ref{scaling2}), i.e., of the existence of a critical value $W_{*c}$ such that the system is delocalized (and ergodic) at $W_* < W_{*c}$ and localized at $W_* > W_{*c}$. 

The reason for the drift of the crossing point to the right in the RRG model was explained in detail in Ref.~\cite{tikhonov2016anderson}, and we briefly recall it here.  The critical point on the RRG is of ``nearly localized'' nature and in particular is characterized by the Poisson level statistics. Thus, the crossing point moves towards the Poisson value of $r$, corresponding to underestimation of $W_*$ from the statistics of a finite-size system. An equivalent way to say this is that in the delocalized phase near the transition point (which corresponds in the present notations to $W_*$ slightly below $W_{*c}$), the spectral statistics shows a non-monotonous behavior as a function of $L$, first approaching the Poisson value $r_P$ and only then starting to grow towards the ergodic value $r_{\rm WD}$. This upturn in the $L$ dependence (or, equivalently, the position of the crossing point between curves with subsequent values of $L$) takes place at the correlation length that scales as $(W_{*c} - W_*)^{-\nu_d}$. The index $\nu_d$ on the delocalized side of the transition on RRG was found to be $\nu_d = 1/2$ \cite{mirlin1991universality,fyodorov1991localization,fyodorov1992novel,tikhonov2016anderson,garcia-mata17}.
\begin{figure*}[ht!]
\centering
\includegraphics[width=0.48\textwidth]{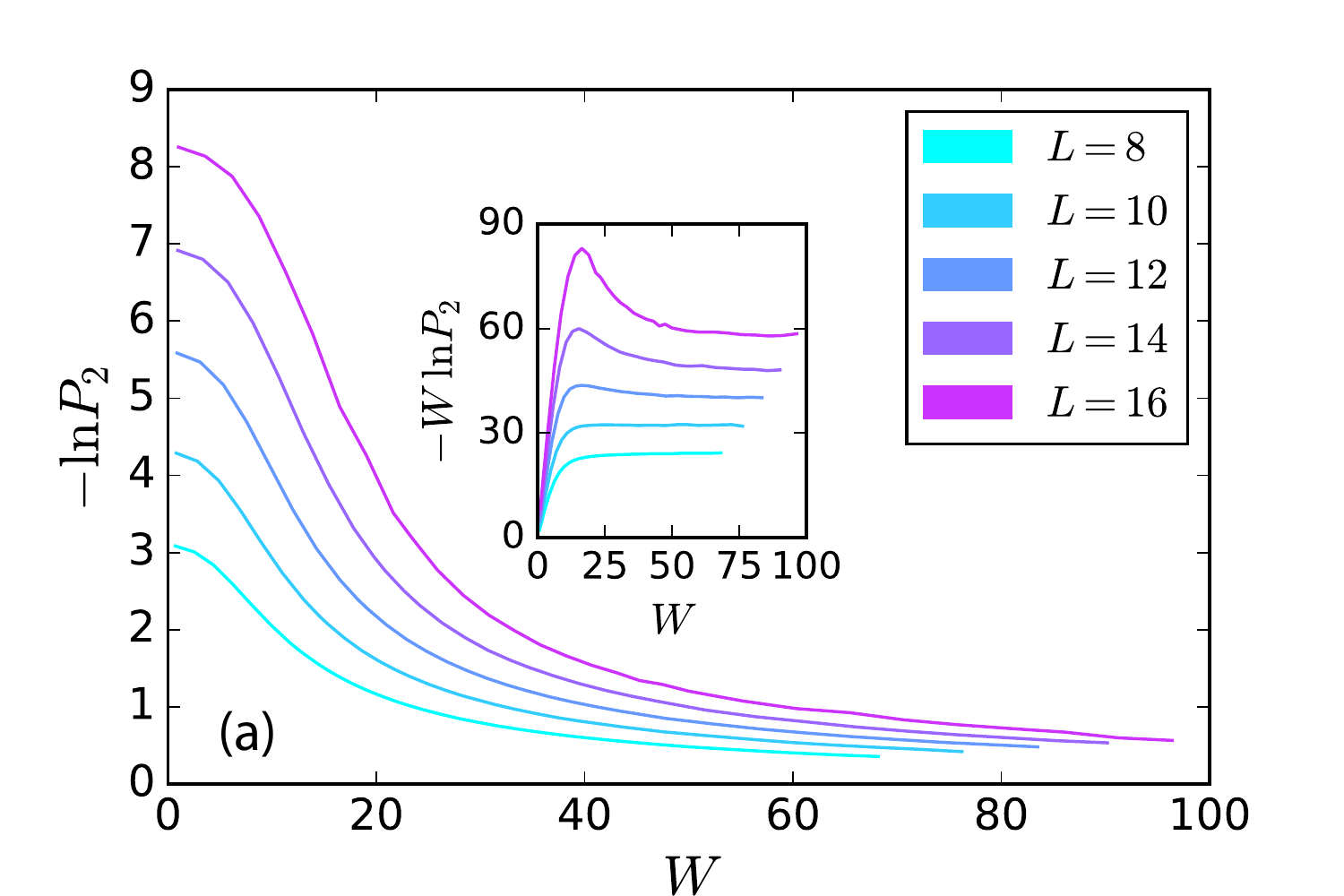}\quad
\includegraphics[width=0.48\textwidth]{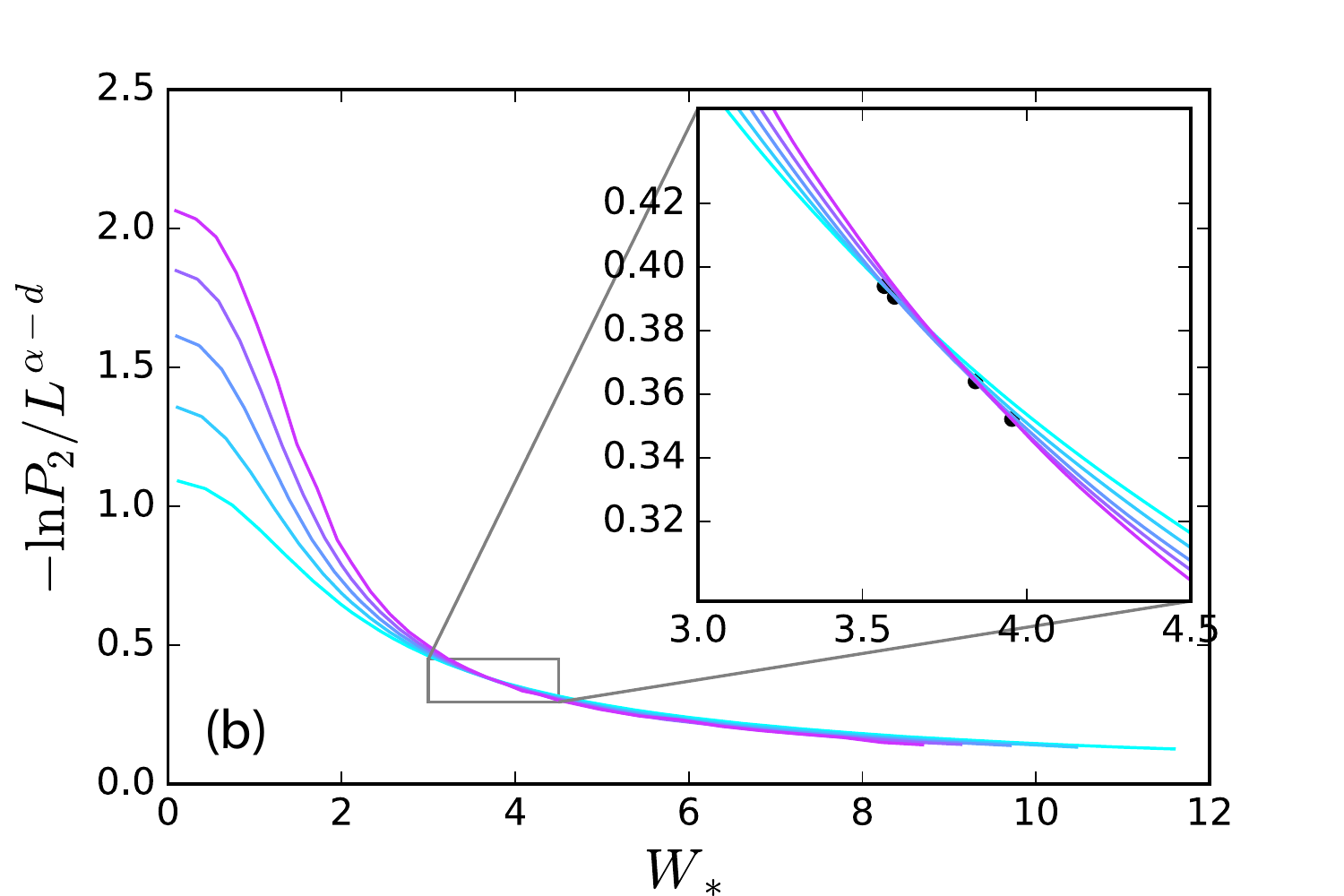}
\caption{Many-body  eigenstate IPR $P_2$  of a 1D spin chain (\ref{long_spin}) with $\alpha=3/2$ as a function of disorder $W$ for various system sizes $L$.
(a) $- \ln P_2$ as a function of $W$ without rescaling. At not too large $W$ the system reaches ergodicity for given system sizes. For large $W$ the system is still in the localized phase for these values of $L$. Inset:  $- \ln P_2$ as a function of $W$ for various $L$.  The behavior given by Eqs.~(\ref{p2-scaling}) and (\ref{nps1}) is manifest at large $W$. (b)  $- \ln P_2 / L^{\alpha-d}$ as a function of the rescaled disorder $W_*$, Eq.~(\ref{W-star}). A slowly drifting crossing point is observed that is expected to converge to the critical value $W_{*c}$ in the limit $L \to \infty$, see Fig.~\ref{fig:crossing}. 
}
\label{fig:ipr}
\end{figure*}

\section{Wavefunction statistics.}
\label{sec:WF}

Let us now turn to properties of many-body eigenfunctions of the Hamiltonian (\ref{long_spin}). 
More specifically, we will characterize the eigenfunctions $\psi^{(j)} \equiv |j\rangle$ by IPR 
\begin{equation}
P_2^{(j)} =\sum_\mu|\psi^{(j)}_\mu|^4 \equiv \sum_\mu |\langle \mu | j\rangle|^4,
\label{P2j}
\end{equation} 
with $\psi_\mu^{(j)} \equiv \langle \mu | j\rangle $ being the wavefunction amplitude on the basis state $\mu$ of the many-body Hilbert space. The basis states $|\mu\rangle$ are eigenstates of all $\hat \sigma_i^z$ operators.  We now analyze the expected behavior of IPR and then turn to comparison to the results of exact diagonalization.

\subsection{Analytical considerations}
\label{sec:ipr-analytical}

We begin with analytical discussion of the IPR scaling in the localized and delocalized phases as well as at the MBL transition.

\subsubsection{Localized phase}
\label{sec:ipr-localized}

We consider first the localized phase, $W_* > W_{*c}$ [which corresponds to $L < L_c(W)$]. In the extreme localization limit, $W_* = \infty$, the eigenstates are identical to the basis states, yielding the largest possible IPR, $P_2=1$.  For the RRG problem, the IPR would remain of the same order,   $P_2 \sim 1$, in the whole localized phase. However, the many-body problem that we are exploring is different from the RRG model in this respect. This is related to the small-scale pseudospins which, while not establishing ergodicity, mix every basis state with a large number of other basis states.
Indeed, according to Eq.~(\ref{rho-ps}), a total number of pseudospins in a system of size $L$ reads
\be
\label{nps}
N_{\rm PS}\sim L^d \frac{t\rho^2}{W}\int_{\rho^{-1/d}}^L \frac{dR}{R}R^{d-\alpha}\sim t\rho^{1+\alpha/d} \frac{L^d}{W}.
\ee 
In the last expression we have assumed $\alpha > d$; in the case $\alpha= d$ an additional logarithmic factor emerges. Setting $\rho = 1$ and $t=1$ (as was sone in Sec.~\ref{sec:num-spectral}), we thus have
\be 
\label{nps1}
N_{\rm PS}\sim \left\{ \begin{array}{cc} 
\displaystyle \frac{L^d}{W}, & \alpha > d; \\[0.3cm]
\displaystyle \frac{L^d \ln L}{W} , & \alpha = d.
\end{array}
\right.
\ee
While these resonances do not lead to delocalization in the considered regime $L<L_c(W)$, their existence is manifest in the scaling of the IPR with the system size. Indeed, each resonance contributes a factor of $\sim 1/2$ to the IPR, thus yielding
\be
- \ln P_2\sim N_{\rm PS},
\label{p2-scaling}
\ee
with $N_{\rm PS}$ given by Eq.~(\ref{nps1}). The emerging scaling of the IPR  looks formally as fractality of eigenstates in the localized phase. Indeed, since the volume of the many-body Hilbert space is 
\be
{\cal N} = 2^{L^d},
\label{hilbert-space-volume}
\ee
Eq.~(\ref{p2-scaling}) can be rewritten (for $\alpha > d$) as $P_2 \sim \cal N^{-\tau}$, with $\tau \sim 1/W$. It is worth mentioning that such fractal scaling of the IPR with $\cal N$ equally applies to the many-body localized phase of a system with short-range interaction \cite{gornyi2017spectral}, as was also observed numerically \cite{luitz2015many}.

\subsubsection{Critical point}

Now we discuss the transition point, $W_* = W_{*c}$ [or, equivalently, $L = L_c(W)$]. In the RRG model, the IPR remains a quantity of order unity also in the localization transition point. In this sense, the RRG  model is different from a $d$-dimensional Anderson transition problem, where IPR has a fractal behavior at criticality. (This is related to the effectively infinite-dimensional character of the RRG model.) In the present situation, the effect of short-scale resonances (see Sec.~\ref{sec:ipr-localized}) will be superimposed on the RRG-type behavior. As a result, the scaling of IPR at the transition point is obtained by setting $L = L_c(W)$ in formulas for the localized phase, Eqs.~(\ref{p2-scaling}) and (\ref{nps1}). This yields the following results:
\be
\label{p2-critical}
-\ln P_2 \sim
\left\{ \begin{array}{cc} 
\displaystyle \frac{L^{\alpha-d}}{\ln L}, & \alpha > d; \\[0.3cm]
1 , & \alpha = d,
\end{array}
\right.
\ee
where $L = L_c(W)$. 

\subsubsection{Delocalized phase}

Finally, we consider the delocalized phase, $W_* < W_{*c}$ [which corresponds to $L > L_c(W)$]. In view of the connection to the RRG problem, we expect that the system becomes ergodic in the large-$L$ limit for a fixed value of $W_*$ smaller than the critical value $W_{*c}$. This corresponds to the IPR proportional to the inverse volume of the Hilbert space  $1/{\cal N}$, i.e.,
\be 
\label{p2-deloc}
- \ln P_2 \simeq L^d \ln 2. 
\ee

\subsection{Numerical results}
\label{ipr-numerical}

Numerical results for the many-body IPR for the 1D model with $\alpha = 3/2$ are shown in Fig.~\ref{fig:ipr}. At not too strong disorder $W$,
the system reaches ergodic behavior  (\ref{p2-deloc}) already for the system sizes that can be treated by exact diagonalization, see Fig.~\ref{fig:ipr}a. 
On the other hand, for large $W$ the system is still in the localized phase for these values of $L$. The inset of Fig.~\ref{fig:ipr}a confirms the behavior predicted for the localized phase,  Eqs.~(\ref{p2-scaling}) and (\ref{nps1}). In Fig.~\ref{fig:ipr}b we plot the rescaled logarithm of the IPR, $- \ln P_2 / L^{\alpha-d}$, as a function of the rescaled disorder $W_*$, Eq.~(\ref{W-star}).  The rescaling along the $y$ axis is chosen in such a way that the corresponding quantity increases with $L$ in the delocalized phase ($W_* < W_{*c}$) and decreases in the localized phase  and in the critical point ($W_* \ge W_{*c}$) according to our analytical predictions, see Sec.~\ref{sec:ipr-analytical}. Therefore, a crossing point drifting to the right and converging to $W_{*c}$ is expected, in analogy with Fig.~\ref{fig:r}b for the levels statistics. This is indeed what is observed in Fig.~\ref{fig:ipr}b. 

In Fig.~\ref{fig:crossing} we have combined the results for the position of the drifting crossing points obtained from the analysis of the level statistics (Fig. \ref{fig:r}b) and the eigenfunction IPR (Fig. \ref{fig:ipr}b). Results of both approaches are consistent with each other and allow us to roughly estimate the critical disorder,
$W_{*c} \simeq 4.3$. When performing this extrapolation, we assumed the value of the critical exponent of the correlation length in the delocalized phase, $\nu_d = 1/2$, see Sec.~\ref{sec:phase-diag}. We have also discarded the data for the smallest system size $L=8$, as the corresponding crossing point in Fig.~\ref{fig:r}b is still close to the Wigner-Dyson value and thus too far from the asymptotic ($L \to \infty$) Poisson value. The fact that the size $L=8$ is too small for being taken into account in a quantitative extrapolation to the thermodynamic limit is also clear from the inset of Fig.~\ref{fig:ipr}. While curves at large $L$ show there a maximum at a disorder of order of the critical one $W_c(L)$, there is still no trace of this maximum for $L=8$.

\section{Critical regime}
\label{sec:phase-diag}

In this section, we briefly discuss the phase diagram of Fig.~\ref{fig:phase-diagram} and in particular the expected width of the critical regime around $W_{*c}$.

As has been explained above, the system is expected to be in the delocalized (respectively, localized) phase in the large $L$-limit if $W_{*}$ is smaller (respectively lager) than $W_{*c}$. Let us consider the value of $W_{*}$ close (but not equal) to the critical value $W_{*c}$. Using the approximate mapping to the RRG model, we then expect the emergence of a large correlation length $\zeta$  in the Hilbert space of the problem that diverges at the transition point according to a power law: 
\begin{eqnarray}
& \zeta(W_*) \sim (W_* - W_{*c})^{-\tilde{\nu}_l}, & \ \ \  W_* > W_{*c};  \nonumber \\
& \zeta(W_*) \sim (W_{*c} - W_*)^{-\tilde{\nu}_d}, & \ \ \  W_* < W_{*c},
\label{corr-length}
\end{eqnarray}
where we have allowed for two different exponents $\tilde{\nu}_l$ and $\tilde{\nu}_d$ on the localized and delocalized sides of the transition.

 The length $\zeta$ determines the ``correlation volume'' in the Hilbert space. In the RRG model with connectivity $K$ the correlation volume is $\sim K^\zeta$.  In view of the analysis of Sec.~\ref{sec:scaling}, we will use $K_{\rm eff} \sim L^{\alpha}$ as an effective coordination number. Equating $K_{\rm eff} ^\zeta$ to the Hilbert space volume given by Eq.~(\ref{hilbert-space-volume}), 
we find the condition for the boundary of the critical regime,
\be
\zeta \sim \frac{L^d}{\ln L},
\label{crit-regime}
\ee
where $\zeta$ is given by Eq.~(\ref{corr-length}). This regime around the critical line is shown in Fig.~\ref{fig:phase-diagram}  by gray color.  If we move along a line of fixed $W_{*}$ close to $W_{*c}$ on the phase diagram (dashed lines in Fig.~\ref{fig:phase-diagram}), we will be first in the critical regime but then [when the length $L$ will exceed the one determined by Eq.~(\ref{crit-regime})] we will end up in either the localized or delocalized phase. In this sense, the MBL localization transition at $W_* = W_{*c}$ becomes sharp in the thermodynamic limit $L\to \infty$, in full analogy with conventional localization transitions. 

\begin{figure}[!h]
\includegraphics[width=\linewidth]{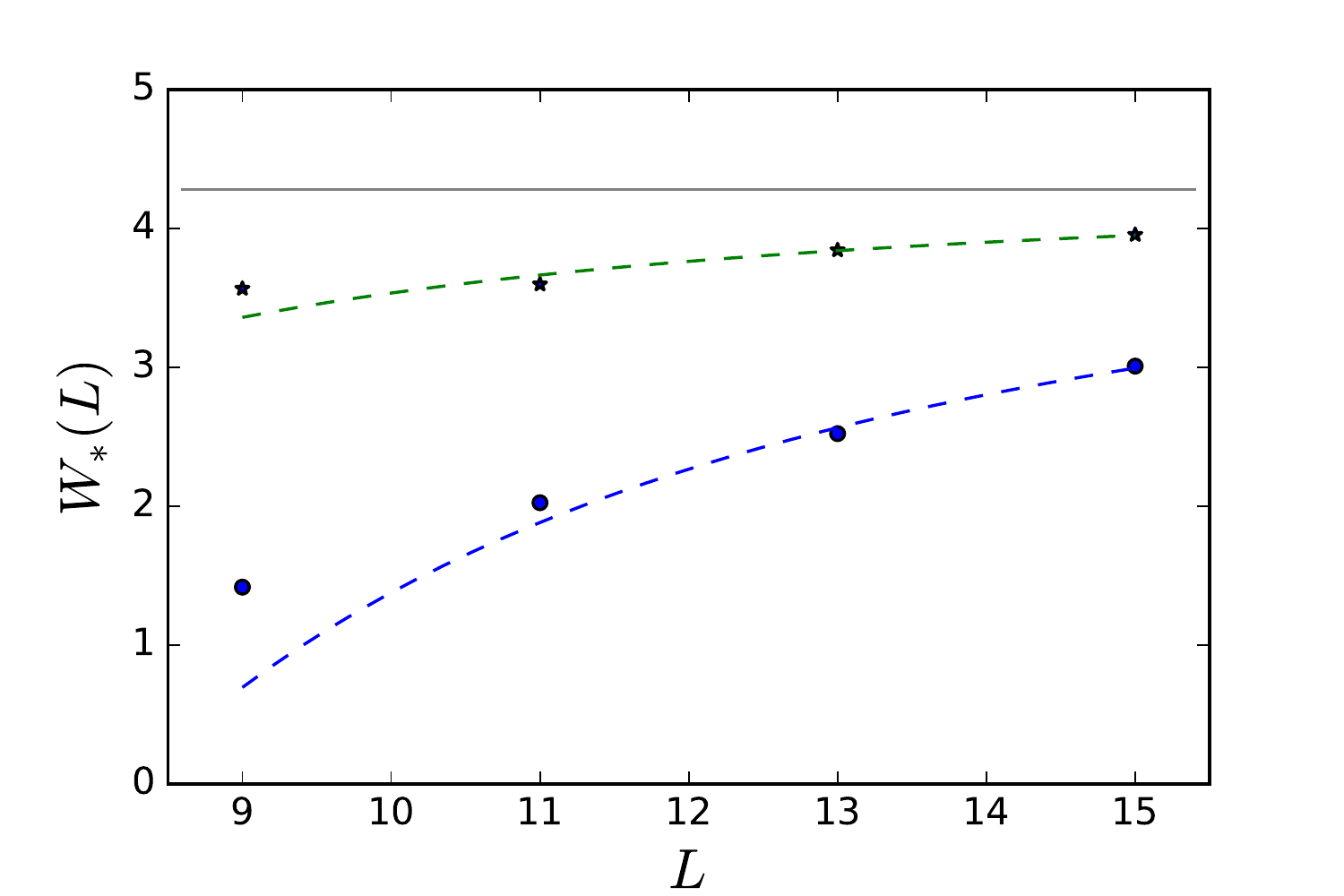}
\caption{Position of the crossing point in $r(W_*)$ curves (Fig. \ref{fig:r}b, circles) and  $\ln P_2(W_*)/L^{\alpha-d}$ (Fig. \ref{fig:ipr}b, stars). Extrapolation of the positions of the crossing point to $L \to \infty$ according to $W_{*}(L)=W_{*c} - {\rm const}\,L^{-2}$ (see text in Sec.~\ref{ipr-numerical} for more detail) renders an estimate for the position of the critical point in the thermodynamic limit: $W_{*c}\simeq 4.3$ (shown by horizontal line).}
\label{fig:crossing}
\end{figure}

Equations (\ref{corr-length}) and (\ref{crit-regime}) give for the finite-size correlation length in the real space $\xi\propto \zeta^{1/d}$, implying critical indices $\nu_{l,d}=\tilde{\nu}_{l,d}/d$ (up to logarithmic corrections).  For the case of the RRG model, $\tilde{\nu}_l = 1$ and $\tilde{\nu}_d = 1/2$ which would yield values of $\nu_{l,d}$ in conflict with Harris criterion $\nu\ge2/d$ (see Ref.~\cite{chandran2015finite} for discussion of Harris criterion for the MBL transition). Apparently, the actual values of $\nu_{l,d}$ describing asymptotic scaling in the vicinity of the transition are different from those suggested by RRG model. The origin of the failure of the mapping to RRG for description of the true critical behavior becomes clear from counting the independent random parameters in both models. Specifically, the number of random variables in the Hamiltonian (\ref{long_spin}) is power-law in the system size, whereas it is exponential in the RRG counterpart. The fluctuations related to finite system size are in fact stronger in the model (\ref{long_spin}) which should lead to larger values of $\nu_{l,d}$. We note, however, that numerical works on MBL transition in 1D systems with short-range interaction yield critical indices $\nu$ in the range $0.5-1$~\cite{kjall2014many,luitz2015many}, also in a strong conflict with Harris criterion. The tentative resolution of this apparent contradiction is that the true asymptotic behavior shows up only in quite large systems $L\gtrsim 500-5000$~\cite{chandran2015finite}. In view of this, we use RRG critical index $\nu_d=1/2$ while estimating the critical disorder $W_{*c}$ from finite size data in Fig.~\ref{fig:crossing}.

\section{Summary and outlook}
\label{s5}

To summarize, we have analyzed the many-body delocalization transition in systems with long-range interactions decaying with a distance according to a power law $1/r^\alpha$ with $d \le \alpha < 2d$, where $d$ is the spatial dimensionality.  We have argued 
for similarity between this problem and that of Anderson localization on RRG  and found the scaling for the critical disorder with the system size: $W_c(L)\propto L^{2d-\alpha} \ln L$. In the large-$L$ limit, the system exhibits a sharp MBL transition as a function of the reduced disorder $W_*$ given by Eq.~(\ref{W-star}). 
We have also studied the IPR of the many-body wavefunctions and demonstrated their fractal behavior in the localized phase as well as at the critical point. 

We have complemented the analytical considerations by exact-diagonalization numerical study of 1D chain with $\alpha=3/2$. Specifically, we have studied the energy level statistics as well the many-body eigenfunction statistics (IPR). These results confirm the analytical expectations for the scaling parameter $W_*$ controlling the MBL transition, and we have numerically estimated the transition point $W_{*c}$.  Further, the numerical results support the expected behavior in the localized and delocalized phases.  In particular, in the localized phase, $W_* > W_{*c}$ and at criticality, $W_* = W_{*c}$, the level statistics evolves with increasing $L$ towards the Poisson form and the IPR has a fractal scaling. On the other hand, in the delocalized phase, $W_* < W_{*c}$, the level and eigenfunction statistics evolve towards  ergodicity at large $L$.   Of course, since system sizes $L$ that can be treated by exact diagonalization are not too large, a quite broad window about the critical value $W_{*c}$ still belongs to the critical regime for such $L$. 

Let us stress that the approximate mapping to RRG (in the many-body Hilbert space) is not at all trivial. Indeed, as has been mentioned in Sec.~\ref{sec:phase-diag}, the Hamiltonian (\ref{long_spin}) depends on $\sim L^{2d}$ random variables, whereas on-site energies of RRG are $\sim 2^{L^d}$ random variables. This implies strong correlations between the energies in the actual many-body problem which are not present in the RRG model. If the second term in the Eq. (\ref{bar-epsilon}) is neglected, these correlations would prevent the system from ergodization even in the presence of pseudospins of the size $\sim L$. It is the spectral diffusion that  strongly reduces the effect of these correlations, restoring the similarity to RRG.

On the experimental side, our results are relevant to a variety of realizations of disordered many-body systems with power-law interactions, see references in Sec.~\ref{s1}. We hope that the future experimental work will allow to study the scaling of the position of MBL transitions in these systems with the system size as well as the physical properties around the transition. It is well known that experimental investigations of MBL transitions represent a highly non-trivial and challenging task. However, recent years have witnessed impressive advances in this direction. This includes also measurements of statistical properties of many-body energy levels and eigenstates that were considered in our paper. In particular, the statistics for many-body energy levels was studied experimentally in a system of superconducting qubits in Ref.~\cite{roushan17}. A crossover from the Wigner-Dyson to the Poisson statistics was observed, which serves as a hallmark of the MBL transition. A complete experimental characterization of many-body eigenstates via expansion over the basis states  is also possible for not too large systems.
In the experiment of Ref.~\cite{haffner2005scalable} such a full quantum state tomography has been carried out for a many-body state of eight trapped ions. As the number of basis states grows exponentially with the system size $L$, this method is restricted to relatively small systems, which are at the same time available to exact numerical diagonalization. 

Remarkably, the IPR of many-body states can be experimentally studied even for much larger systems. Indeed, recent experiments demonstrated coherent quantum evolution in many-body systems of  53 trapped ions \cite{zhang2017observation} and 51 atoms \cite{bernien2017probing}, which can be described by Hamiltonians of coupled spins 1/2.   In these setups,  measuring all coefficients in the expansion of a many-body state over the complete basis is impossible in view of a huge number of the basis states, ${\cal N} \sim 2^{50} \sim 10^{15}$. However, a
projection of the quantum state to  certain selected basis states has been measured via single-shot state detection. In this way, one can prepare the system initially in a given basis state $|\psi(0)\rangle = |\mu\rangle$ and then measure the 
probability of return of the quantum state (which develops according to the full many-body Hamiltonian) to the same basis state after a time $t$:
\begin{equation}
C(t)=\left|\left<\psi(t)|\psi(0)\right>\right|^2.
\end{equation}
Expanding $|\psi(t)\rangle$ in exact eigenstates $|i\rangle$, one gets
\be
C(t) = \sum_{jk}e^{-i(E_j-E_k)t} |\langle j| \mu\rangle|^2 |\langle k| \mu\rangle|^2.
\ee
In the long-time limit, we can discard oscillatory terms, which yields the IPR in the many-body Hilbert space:
\be
C(t \to \infty) = \sum_j |\langle j| \mu\rangle|^4 \equiv P_2^{(\mu)}.
\label{P2mu}
\ee
In fact, there is a slight difference in the definition of IPR between Eqs.~(\ref{P2j}) and (\ref{P2mu}): the former one corresponds to expansion of a given exact many-body state over the Fock-space basis state, while the latter is a dual quantity. This difference is not essential, however, when the average value is calculated.

Therefore, experimental determination of IPR  $P_2$ can be carried out via single-shot measurements of the many-body return probability $C(t)$.  This is feasible, as long as $P_2$ is not too small, since one has to perform the measurement $\sim P_2^{-1}$ times.   We have shown, however, in this paper that the critical point of the MBL transition in the problem with power-law interaction is similar to the localized phase and, in particular, the IPR at criticality is not too small, see Eq.(\ref{p2-critical}). This makes an experimental investigation of IPR in the critical point as well as in its vicinity feasible even for quite large systems. As a model example, let us consider a 1D system with interaction  exponent $\alpha=3/2$ (as studied numerically in our work). Let's assume that we perform the measurement of the order of $10^4$ times (as in Refs.~\cite{zhang2017observation,bernien2017probing}), i.e. can measure $P_2$ as long as it is larger than $\sim 10^{-4}$. Using Eq.~(\ref{p2-critical}), we estimate that $P_2$ at criticality can then be measured for systems with length up to $L \sim 7000$, despite the fact that the total volume of the Hilbert space ${\cal N} = 2^L$ for such systems is astronomically large, ${\cal N} \sim 2^{7000} \sim 10^{2000}$. As a second example, we consider a 2D system with dipole-dipole interaction, $\alpha =3$. Repeating the same estimate, we find that $P_2$ at the MBL transition can be determined for system sizes up to $L \sim 30$, i.e. with spatial volume up to $L^2 \sim 1000$. (Again, the total volume of the Hilbert space for such systems is enormous, ${\cal N} = 2^{L^2} \sim 10^{300}$.) Our estimates thus show that the study of IPR around the MBL transition in systems with long-range interaction is in principle possible in rather large systems. The fact that systems with $\approx 50$ qubits have already been experimentally implemented as well as the rapid progress in this field allow us to hope that such measurements can be carried out in a not too far future.  In addition to analysis of the eigenfunction statistics, the MBL transition in system with power-law interaction can be experimentally detected also by investigation of other physical observables such as, e.g., the spin relaxation or the energy transport.

We also hope that our work will pave the way for more detailed theoretical studies of the MBL transitions. The mechanism of the  transition in the model addressed in this work appears to be somewhat simpler than in quantum dots and in spatially extended systems. This manifests itself in a particularly close connection with the Anderson transition on RRG. Nevertheless, this connection is not rigorously understood at this stage yet. This, in particular, applies to the critical exponents $\nu_d$ and $\nu_l$, see Sec.~\ref{sec:phase-diag}. 

\begin{figure}[!h]
\includegraphics[width=\linewidth]{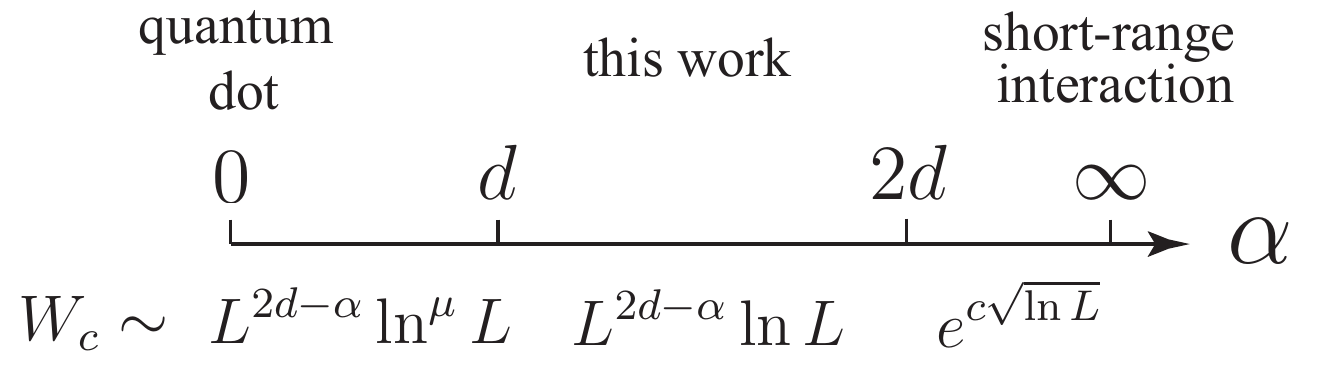}
\caption{Evolution of the critical disorder $W_c$ of the ergodization transition with the power-law-interaction exponent $\alpha$. 
The range $d \le \alpha < 2d$ is considered in the present work, and the corresponding critical disorder is given by Eq.~(\ref{scaling2}). Extreme cases are the limits of infinite-range interaction ($\alpha=0$, quantum dot) and of short-range interaction ($\alpha = \infty$). In the range $0 \le \alpha < d$ the mechanism of ergodization is analogous to that in the quantum dot ($\alpha=0$) model discussed in Ref.~\cite{gornyi2017spectral}. In the range $2d \le \alpha < \infty$, the delocalization is expected to take place due to rare ergodic spots \cite{roeck17,thiery2017microscopically}. See text for more detail.
 }
\label{fig:various-alpha}
\end{figure}

It is worth mentioning another interacting model with power-law decaying terms, which has been studied in Ref. \cite{PhysRevE.85.050102}. This is PRBM model (see Sec. \ref{s2}), supplemented by short-range interaction. As numerical results suggest, the critical value of the exponent $\alpha$ for this model is larger than non-interacting critical value $\alpha=1$.
It might be interesting to study the delocalization transition in that model as well as possible relations with the model studied in the present work.

Before closing, let us briefly discuss what happens with the MBL transition for other values of the power-law interaction exponent $\alpha$. We recall that in this work we focused on the range $d \le \alpha < 2d$.  In general, one can vary $\alpha$ between 0 and $\infty$. The limit $\alpha=0$ corresponds to an infinite-range interaction, i.e., to a spin quantum dot model where all interaction matrix elements are of the same order. The opposite limit $\alpha=\infty$ corresponds to a model with short-range interaction. The mechanism and the scaling of the ergodization transition (delocalization in the many-body Hilbert space) in the spin quantum dot model was considered in Ref.~\cite{gornyi2017spectral}. A direct extension of the analysis in that work to the range $0 \le \alpha < d$ yields $W_c \sim L^{2d-\alpha} \ln^\mu L$, where the index $\mu \le 1$ (that can depend on $\alpha$) remains to be found. A lower bound on $\mu$ can be found using the approach of Ref.~\cite{gornyi2017spectral}. The mechanisms of ergodization for  $0 \le \alpha < d$ and $d \le \alpha < 2d$ bear some similarity: in both cases the transition takes place, up to logarithmic factors, when first pseudospins of size $\sim L$ emerge. However, the spectral diffusion---which plays a key role for establishing ergodicity---is more efficient for $d \le \alpha < 2d$ than for $0 \le \alpha < d$, which makes the analysis for $0 \le \alpha < d$ and in particular the determination of the exponent $\mu$ more difficult. 
For faster decaying interaction, $2d < \alpha < \infty$, the delocalization mechanism considered in the present work is not operative any more.
There exists, however, another delocalization mechanism---the one related to rare ergodic spots \cite{roeck17,thiery2017microscopically}. (We have not discussed it above since in the regime $\alpha < 2d$ the delocalization mechanism considered in our work is much more efficient.) 
That mechanism is expected to establish ergodization in the large-$L$ limit (at fixed $W$) for systems with $\alpha \ge 2d$. To estimate the corresponding $W_c(L)$, we note that delocalization via the mechanism of Refs.~\onlinecite{roeck17,thiery2017microscopically} in the power-law interaction model will happen if an ergodic spot of volume $V \sim \ln W$ emerges. This will be the case if all random energies $\epsilon_i$ within this spot are of order of unity, which yields a probability of such a rare event $\sim \exp( - {\rm const}\, \ln^2W)$.  Thus, the critical length $L_c(W)$ can be estimated as $L_c \sim \exp( {\rm const}\, \ln^2W)$. Equivalently, we get an estimate for $L$-dependent critical disorder, $W_c \sim \exp( {\rm const}\, \ln^{1/2}L)$. The estimated behavior of $W_c(L)$ in the full range of $\alpha$ is summarized in Fig.~\ref{fig:various-alpha}.

\section{Acknowledgments}
We thank A. Burin and A. Garcia-Garcia for useful comments. This work was supported by Russian Science Foundation under Grant No. 14-42-00044.
KT acknowledges support by Alexander von Humboldt Foundation.
\bibliography{spins}

\end{document}